
%
%
  \font\twelverm=cmr10 scaled 1200       \font\twelvei=cmmi10 scaled 1200
  \font\twelvesy=cmsy10 scaled 1200      \font\twelveex=cmex10 scaled 1200
  \font\twelvebf=cmbx10 scaled 1200      \font\twelvesl=cmsl10 scaled 1200
  \font\twelvett=cmtt10 scaled 1200      \font\twelveit=cmti10 scaled 1200
  \font\twelvemib=cmmib10 scaled 1200
  \font\elevenmib=cmmib10 scaled 1095
  \font\tenmib=cmmib10
  \font\eightmib=cmmib10 scaled 800
  
\font\elevenrm=cmr10 scaled 1095    \font\eleveni=cmmi10 scaled 1095
\font\elevensy=cmsy10 scaled 1095

%
%

\font\seventeeni=cmmi10 scaled \magstep3

\font\seventeensy=cmsy10 scaled \magstep3

\font\seventeenmib=cmmib10 scaled \magstep3

\newfam\cpfam%



\skewchar\eleveni='177   \skewchar\elevensy='60
\skewchar\elevenmib='177  \skewchar\seventeensy='60
\skewchar\seventeenmib='177
\skewchar\seventeeni='177

\newfam\mibfam%


  \skewchar\twelvei='177   \skewchar\twelvesy='60
  \skewchar\twelvemib='177
%
%
\def\twelvepoint{\normalbaselineskip=12.4pt
  \abovedisplayskip 12.4pt plus 3pt minus 9pt
  \belowdisplayskip 12.4pt plus 3pt minus 9pt
  \abovedisplayshortskip 0pt plus 3pt
  \belowdisplayshortskip 7.2pt plus 3pt minus 4pt
  \smallskipamount=3.6pt plus 1.2pt minus 1.2pt
  \medskipamount=7.2pt plus 2.4pt minus 2.4pt
  \bigskipamount=14.4pt plus 4.8pt minus 4.8pt
  \def\rm{\fam0\twelverm}          \def\it{\fam\itfam\twelveit}%
  \def\sl{\fam\slfam\twelvesl}     \def\bf{\fam\bffam\twelvebf}%
  \def\mit{\fam 1}                 \def\cal{\fam 2}%
  \def\tt{\twelvett}%
  \def\mib{\fam\mibfam\twelvemib}%

  \textfont0=\twelverm   \scriptfont0=\tenrm     \scriptscriptfont0=\sevenrm
  \textfont1=\twelvei    \scriptfont1=\teni      \scriptscriptfont1=\seveni
  \textfont2=\twelvesy   \scriptfont2=\tensy     \scriptscriptfont2=\sevensy
  \textfont3=\twelveex   \scriptfont3=\twelveex  \scriptscriptfont3=\twelveex
  \textfont\itfam=\twelveit
  \textfont\slfam=\twelvesl
  \textfont\bffam=\twelvebf
  \textfont\mibfam=\twelvemib       \scriptfont\mibfam=\tenmib
                                             \scriptscriptfont\mibfam=\eightmib

  \def\xrm{\textfont0=\twelverm\scriptfont0=\tenrm
      \scriptscriptfont0=\sevenrm\rm}
\normalbaselines\rm}


\mathchardef\alpha="710B
\mathchardef\beta="710C
\mathchardef\gamma="710D
\mathchardef\delta="710E
\mathchardef\epsilon="710F
\mathchardef\zeta="7110
\mathchardef\eta="7111
\mathchardef\theta="7112
\mathchardef\kappa="7114
\mathchardef\lambda="7115
\mathchardef\mu="7116
\mathchardef\nu="7117
\mathchardef\xi="7118
\mathchardef\pi="7119
\mathchardef\rho="711A
\mathchardef\sigma="711B
\mathchardef\tau="711C
\mathchardef\phi="711E
\mathchardef\chi="711F
\mathchardef\psi="7120
\mathchardef\omega="7121
\mathchardef\varepsilon="7122
\mathchardef\vartheta="7123
\mathchardef\varrho="7125
\mathchardef\varphi="7127

\def\physgreek{
\mathchardef\Gamma="7100
\mathchardef\Delta="7101
\mathchardef\Theta="7102
\mathchardef\Lambda="7103
\mathchardef\Xi="7104
\mathchardef\Pi="7105
\mathchardef\Sigma="7106
\mathchardef\Upsilon="7107
\mathchardef\Phi="7108
\mathchardef\Psi="7109
\mathchardef\Omega="710A}


\def\beginlinemode{\endmode
  \begingroup\parskip=0pt \obeylines\def\\{\par}\def\endmode{\par\endgroup}}
\def\beginparmode{\endmode
  \begingroup \def\endmode{\par\endgroup}}
\let\endmode=\par
{\obeylines\gdef\
{}}
\def\singlespace{\baselineskip=\normalbaselineskip}

\def\oneandahalfspace{\baselineskip=\normalbaselineskip
  \multiply\baselineskip by 3 \divide\baselineskip by 2}
\def\doublespace{\baselineskip=\normalbaselineskip \multiply\baselineskip by 2}

\nopagenumbers
\newcount\firstpageno
\firstpageno=2
\headline={\ifnum\pageno<\firstpageno{\hfil}\else{\hfil\elevenrm\folio\hfil}\fi}
\let\rawfootnote=\footnote             
\def\footnote#1#2{{\singlespace\parindent=0pt
\rawfootnote{#1}{#2}}}
\def\raggedcenter{\leftskip=4em plus 12em \rightskip=\leftskip
  \parindent=0pt \parfillskip=0pt \spaceskip=.3333em \xspaceskip=.5em
  \pretolerance=9999 \tolerance=9999
  \hyphenpenalty=9999 \exhyphenpenalty=9999 }
\def\dateline{\rightline{\ifcase\month\or
  January\or February\or March\or April\or May\or June\or
  July\or August\or September\or October\or November\or December\fi
  \space\number\year}}
\def\received{\vskip 3pt plus 0.2fill
 \centerline{\sl (Received\space\ifcase\month\or
  January\or February\or March\or April\or May\or June\or
  July\or August\or September\or October\or November\or December\fi
  \qquad, \number\year)}}


\hsize=6.5truein
\hoffset=0.0truein
\vsize=8.9truein
\voffset=0truein
\hfuzz=0.1pt
\vfuzz=0.1pt
\parskip=\medskipamount
\overfullrule=0pt      



\def\title                     
  {\null\vskip 3pt plus 0.1fill
   \beginlinemode \doublespace \raggedcenter \bf}

\def\author                    
  {\vskip 6pt plus 0.2fill \beginlinemode
   \singlespace \raggedcenter}

\def\affil        
  {\vskip 6pt plus 0.1fill \beginlinemode
   \oneandahalfspace \raggedcenter \it}

\def\abstract                  
  {\vskip 6pt plus 0.3fill \beginparmode
   \doublespace \narrower }

\def\summary                   
  {\vskip 3pt plus 0.3fill \beginparmode
   \doublespace \narrower SUMMARY: }

\def\pacs#1
  {\vskip 3pt plus 0.2fill PACS numbers: #1}

\def\endtitlepage              
  {\endpage                    
   \body}

\def\body                      
  {\beginparmode}              

\def\head#1{                   
  \filbreak\vskip 0.5truein    
  {\immediate\write16{#1}
   \raggedcenter \uppercase{#1}\par}
   \nobreak\vskip 0.25truein\nobreak}

%
%

%
\def\inlinerefs{
  \gdef\refto##1{ [##1]}                
\gdef\refis##1{\indent\hbox to 0pt{\hss##1.~}} 
\gdef\journal##1, ##2, ##3, 1##4##5##6{ 
    {\sl ##1~}{\bf ##2}, ##3 (1##4##5##6)}}    
\def\keywords#1
  {\vskip 3pt plus 0.2fill Keywords: #1}
\gdef\figis#1{\indent\hbox to 0pt{\hss#1.~}} 

\def\figurecaptions     
  {\head{Figure Captions}    
   \beginparmode
   \interlinepenalty=10000
   \frenchspacing \parindent=0pt \leftskip=1truecm
   \parskip=8pt plus 3pt \everypar{\hangindent=\parindent}}

%
%
\def\refto#1{$^{#1}$}          

\def\references       
  {\head{References}           
   \beginparmode
   \frenchspacing \parindent=0pt \leftskip=1truecm
   \interlinepenalty=10000
   \parskip=8pt plus 3pt \everypar{\hangindent=\parindent}}

\gdef\refis#1{\indent\hbox to 0pt{\hss#1.~}} 

\gdef\journal#1, #2, #3, 1#4#5#6{              
    {\sl #1~}{\bf #2}, #3 (1#4#5#6)}          

\def\refstylenp{               
  \gdef\refto##1{ [##1]}                               
  \gdef\refis##1{\indent\hbox to 0pt{\hss##1)~}}      
  \gdef\journal##1, ##2, ##3, ##4 {                    
     {\sl ##1~}{\bf ##2~}(##3) ##4 }}

\def\refstyleprnp{             
  \gdef\refto##1{ [##1]}                               
  \gdef\refis##1{\indent\hbox to 0pt{\hss##1)~}}      
  \gdef\journal##1, ##2, ##3, 1##4##5##6{              
    {\sl ##1~}{\bf ##2~}(1##4##5##6) ##3}}

\def\prb{\journal Phys. Rev. B, }

\def\prl{\journal Phys. Rev. Lett., }

\def\jpsj{\journal J. Phys. Soc. Japan., }

\def\endreferences{\body}

%
%

\def\endpage                   
  {\vfill\eject}

\def\endpaper                  
  {\endmode\vfill\supereject}

\def\endit
  {\endpaper\end}


\def\ref#1{Ref.[#1]}                   
\def\Ref#1{Ref.[#1]}                   

\def\Equation#1{Equation [#1]}         
\def\Equations#1{Equations [#1]}       
\def\Eq#1{Eq. (#1)}                     
\def\eq#1{Eq. (#1)}                     
\def\Eqs#1{Eqs. (#1)}                   
\def\eqs#1{Eqs. (#1)}                   
\def\frac#1#2{{\textstyle{{\strut #1} \over{\strut #2}}}}

\def\sla{\raise.15ex\hbox{$/$}\kern-.57em}
\def\leaderfill{\leaders\hbox to 1em{\hss.\hss}\hfill}
\def\twiddle{\lower.9ex\rlap{$\kern-.1em\scriptstyle\sim$}}
\def\bigtwiddle{\lower1.ex\rlap{$\sim$}}
\def\gtwid{\mathrel{\raise.3ex\hbox{$>$\kern-.75em\lower1ex\hbox{$\sim$}}}}
\def\ltwid{\mathrel{\raise.3ex\hbox{$<$\kern-.75em\lower1ex\hbox{$\sim$}}}}
\def\square{\kern1pt\vbox{\hrule height 1.2pt\hbox{\vrule width 1.2pt\hskip 3pt
   \vbox{\vskip 6pt}\hskip 3pt\vrule width 0.6pt}\hrule height 0.6pt}\kern1pt}

%

%

%

%
\physgreek
%

\def\dsl{\raise.15ex\hbox{$/$}\kern-.57em\hbox{$\partial$}}
\def\nsl{\raise.15ex\hbox{$/$}\kern-.57em\hbox{$\nabla$}}
\def\gtwid{\,{\raise.3ex\hbox{$>$\kern-.75em\lower1ex\hbox{$\sim$}}}\,}
\def\ltwid{\,{\raise.3ex\hbox{$<$\kern-.75em\lower1ex\hbox{$\sim$}}}\,}
\def\undr{\raise.3ex\hbox{$\sim$\kern-.75em\lower1ex\hbox{$|\vec
x|\to\infty$}}}

\def\[{\left [}
\def\]{\right ]}
\def\({\left (}
\def\){\right )}







\def\and{a^{\phantom\dagger}}

%
\def\id{\raise.72ex\hbox{$-$}\kern-.85em\hbox{$d$}\,}

\catcode`@=11
\newcount\r@fcount \r@fcount=0
\newcount\r@fcurr
\immediate\newwrite\reffile
\newif\ifr@ffile\r@ffilefalse
\def\w@rnwrite#1{\ifr@ffile\immediate\write\reffile{#1}\fi\message{#1}}

\def\writer@f#1>>{}
\def\referencefile{
  \r@ffiletrue\immediate\openout\reffile=\jobname.ref%
  \def\writer@f##1>>{\ifr@ffile\immediate\write\reffile%
    {\noexpand\refis{##1} = \csname r@fnum##1\endcsname = %
     \expandafter\expandafter\expandafter\strip@t\expandafter%
     \meaning\csname r@ftext\csname r@fnum##1\endcsname\endcsname}\fi}%
  \def\strip@t##1>>{}}

\def\citeall#1{\xdef#1##1{#1{\noexpand\cite{##1}}}}
\def\cite#1{\each@rg\citer@nge{#1}}	

\def\each@rg#1#2{{\let\thecsname=#1\expandafter\first@rg#2,\end,}}
\def\first@rg#1,{\thecsname{#1}\apply@rg}	
\def\apply@rg#1,{\ifx\end#1\let\next=\relax
\else,\thecsname{#1}\let\next=\apply@rg\fi\next}

\def\citer@nge#1{\citedor@nge#1-\end-}	
\def\citer@ngeat#1\end-{#1}
\def\citedor@nge#1-#2-{\ifx\end#2\r@featspace#1 
  \else\citel@@p{#1}{#2}\citer@ngeat\fi}	
\def\citel@@p#1#2{\ifnum#1>#2{\errmessage{Reference range #1-#2\space is bad.}%
    \errhelp{If you cite a series of references by the notation M-N, then M and
    N must be integers, and N must be greater than or equal to M.}}\else%
 {\count0=#1\count1=#2\advance\count1
by1\relax\expandafter\r@fcite\the\count0,%
  \loop\advance\count0 by1\relax
    \ifnum\count0<\count1,\expandafter\r@fcite\the\count0,%
  \repeat}\fi}

\def\r@featspace#1#2 {\r@fcite#1#2,}	
\def\r@fcite#1,{\ifuncit@d{#1}
    \newr@f{#1}%
    \expandafter\gdef\csname r@ftext\number\r@fcount\endcsname%
                     {\message{Reference #1 to be supplied.}%
                      \writer@f#1>>#1 to be supplied.\par}%
 \fi%
 \csname r@fnum#1\endcsname}
\def\ifuncit@d#1{\expandafter\ifx\csname r@fnum#1\endcsname\relax}%
\def\newr@f#1{\global\advance\r@fcount by1%
    \expandafter\xdef\csname r@fnum#1\endcsname{\number\r@fcount}}

\let\r@fis=\refis			
\def\refis#1#2#3\par{\ifuncit@d{#1}
   \newr@f{#1}%
   \w@rnwrite{Reference #1=\number\r@fcount\space is not cited up to now.}\fi%
  \expandafter\gdef\csname r@ftext\csname r@fnum#1\endcsname\endcsname%
  {\writer@f#1>>#2#3\par}}

\def\ignoreuncited{
   \def\refis##1##2##3\par{\ifuncit@d{##1}%
     \else\expandafter\gdef\csname r@ftext\csname
r@fnum##1\endcsname\endcsname%
     {\writer@f##1>>##2##3\par}\fi}}

\def\r@ferr{\endreferences\errmessage{I was expecting to see
\noexpand\endreferences before now;  I have inserted it here.}}
\let\r@ferences=\references
\def\references{\r@ferences\def\endmode{\r@ferr\par\endgroup}}

\let\endr@ferences=\endreferences
\def\endreferences{\r@fcurr=0
  {\loop\ifnum\r@fcurr<\r@fcount
    \advance\r@fcurr by 1\relax\expandafter\r@fis\expandafter{\number\r@fcurr}%
    \csname r@ftext\number\r@fcurr\endcsname%
  \repeat}\gdef\r@ferr{}\endr@ferences}


\let\r@fend=\endpaper\gdef\endpaper{\ifr@ffile
\immediate\write16{Cross References written on []\jobname.REF.}\fi\r@fend}

\catcode`@=12

\citeall\refto		
\citeall\ref		%
\citeall\Ref		%

\catcode`@=11
\newcount\tagnumber\tagnumber=0

\immediate\newwrite\eqnfile
\newif\if@qnfile\@qnfilefalse
\def\write@qn#1{}
\def\writenew@qn#1{}
\def\w@rnwrite#1{\write@qn{#1}\message{#1}}
\def\@rrwrite#1{\write@qn{#1}\errmessage{#1}}

\def\taghead#1{\gdef\t@ghead{#1}\global\tagnumber=0}
\def\t@ghead{}

\expandafter\def\csname @qnnum-3\endcsname
  {{\t@ghead\advance\tagnumber by -3\relax\number\tagnumber}}
\expandafter\def\csname @qnnum-2\endcsname
  {{\t@ghead\advance\tagnumber by -2\relax\number\tagnumber}}
\expandafter\def\csname @qnnum-1\endcsname
  {{\t@ghead\advance\tagnumber by -1\relax\number\tagnumber}}
\expandafter\def\csname @qnnum0\endcsname
  {\t@ghead\number\tagnumber}
\expandafter\def\csname @qnnum+1\endcsname
  {{\t@ghead\advance\tagnumber by 1\relax\number\tagnumber}}
\expandafter\def\csname @qnnum+2\endcsname
  {{\t@ghead\advance\tagnumber by 2\relax\number\tagnumber}}
\expandafter\def\csname @qnnum+3\endcsname
  {{\t@ghead\advance\tagnumber by 3\relax\number\tagnumber}}

\def\equationfile{%
  \@qnfiletrue\immediate\openout\eqnfile=\jobname.eqn%
  \def\write@qn##1{\if@qnfile\immediate\write\eqnfile{##1}\fi}
  \def\writenew@qn##1{\if@qnfile\immediate\write\eqnfile
    {\noexpand\tag{##1} = (\t@ghead\number\tagnumber)}\fi}
}

\def\callall#1{\xdef#1##1{#1{\noexpand\call{##1}}}}
\def\call#1{\each@rg\callr@nge{#1}}

\def\each@rg#1#2{{\let\thecsname=#1\expandafter\first@rg#2,\end,}}
\def\first@rg#1,{\thecsname{#1}\apply@rg}
\def\apply@rg#1,{\ifx\end#1\let\next=\relax%
\else,\thecsname{#1}\let\next=\apply@rg\fi\next}

\def\callr@nge#1{\calldor@nge#1-\end-}
\def\callr@ngeat#1\end-{#1}
\def\calldor@nge#1-#2-{\ifx\end#2\@qneatspace#1 %
  \else\calll@@p{#1}{#2}\callr@ngeat\fi}
\def\calll@@p#1#2{\ifnum#1>#2{\@rrwrite{Equation range #1-#2\space is bad.}
\errhelp{If you call a series of equations by the notation M-N, then M and
N must be integers, and N must be greater than or equal to M.}}\else%
 {\count0=#1\count1=#2\advance\count1
 by1\relax\expandafter\@qncall\the\count0,%
  \loop\advance\count0 by1\relax%
    \ifnum\count0<\count1,\expandafter\@qncall\the\count0,%
  \repeat}\fi}

\def\@qneatspace#1#2 {\@qncall#1#2,}
\def\@qncall#1,{\ifunc@lled{#1}{\def\next{#1}\ifx\next\empty\else
  \w@rnwrite{Equation number \noexpand\(>>#1<<) has not been defined yet.}
  >>#1<<\fi}\else\csname @qnnum#1\endcsname\fi}

\let\eqnono=\eqno
\def\eqno(#1){\tag#1}
\def\tag#1$${\eqnono(\displayt@g#1 )$$}

\def\aligntag#1\endaligntag
  $${\gdef\tag##1\\{&(##1 )\cr}\eqalignno{#1\\}$$
  \gdef\tag##1$${\eqnono(\displayt@g##1 )$$}}

\def\eqalignno#1{\displ@y \tabskip\centering
  \halign to\displaywidth{\hfil$\displaystyle{##}$\tabskip\z@skip
    &$\displaystyle{{}##}$\hfil\tabskip\centering
    &\llap{$\displayt@gpar##$}\tabskip\z@skip\crcr
    #1\crcr}}

\def\displayt@gpar(#1){(\displayt@g#1 )}

\def\displayt@g#1 {\rm\ifunc@lled{#1}\global\advance\tagnumber by1
        {\def\next{#1}\ifx\next\empty\else\expandafter
        \xdef\csname @qnnum#1\endcsname{\t@ghead\number\tagnumber}\fi}%
  \writenew@qn{#1}\t@ghead\number\tagnumber\else
        {\edef\next{\t@ghead\number\tagnumber}%
        \expandafter\ifx\csname @qnnum#1\endcsname\next\else
        \w@rnwrite{Equation \noexpand\tag{#1} is a duplicate number.}\fi}%
  \csname @qnnum#1\endcsname\fi}

\def\ifunc@lled#1{\expandafter\ifx\csname @qnnum#1\endcsname\relax}

\let\@qnend=\end\gdef\end{\if@qnfile
\immediate\write16{Equation numbers written on []\jobname.EQN.}\fi\@qnend}

\catcode`@=12
\callall\Equation
\callall\Equations
\callall\Eq
\callall\eq
\callall\Eqs
\callall\eqs


\referencefile

\twelvepoint\doublespace

\title{Effect of Matrix Elements on the Pairing Kernel in Exchange Mediated
Superconductors}

\author{M. R. Norman}
\affil
Materials Science Division
Argonne National Laboratory
Argonne, IL  60439

\abstract

A method is derived for calculating the pairing kernel in exchange mediated
superconductors including matrix element effects.  Various
models for the interaction vertex are considered, including spin
exchange, orbital exchange, and quadrupolar exchange.  As an example,
this formalism is applied to $UPt_3$ using relativistic
wavefunctions from a local density band calculation.

\bigskip

\noindent PACS numbers:  74.20.-z, 74.70.Tx

\endtitlepage

Much work has been done over the past decade in trying to understand the
microscopic basis for superconductivity in heavy fermion metals, and more
recently in the copper oxides.  In the heavy fermion case particularly,
a large body of experimental data points to the presence of
a non-trivial order parameter reminiscent of $^3He$.  Because of this,
many theoretists working on this subject, including the author, have
borrowed techniques successful for $^3He$ and applied them to the heavy
fermion problem.  In particular, a favorite mechanism being explored is
pairing due to spin fluctuations.  In the calculations performed, one uses
some model for the dynamic susceptibility and solves a gap equation.  The
solution is determined by the k dependence of the susceptibility and the
Fermi surface\refto{miyake,norman,putika,grewe}.  Similar calculations have
been recently performed on copper oxide materials\refto{MPB}.

The deficiency of this approach is obvious.  It ignores matrix element
effects, that is, the k dependence enters only through the energy dispersion
and
not from the wavefunctions.  In the analogous magnetic problem, the magnetism
would be determined solely by nesting effects.  Given the commensurate nature
of the magnetic order in heavy fermions, nesting is an unlikely determinator
for magnetism, and thus, one would suspect, for superconductivity also.  This
was realized from the beginning of theoretical work on this subject\refto{and}.
Despite this, little work has been done along these lines,
a noticable exception being a paper by Appel and Hertel\refto{appel}, where
a real space approach to pairing in heavy fermions is advocated.

In this paper, following the spirit of Refs. 6 and 7, the author develops a
formalism for calculating the pairing kernel in exchange mediated
superconductors including
relativistic matrix elements determined from band structure calculations.
Various models for the pairing interaction are considered, including spin,
orbital, and quadrupolar exchange.  As an example, this method is applied
to the case of
$UPt_3$, the best studied of the heavy fermion superconductors.

The pairing kernel to be evalulated is of the form $<k,-k|V|k',-k'>$
where V is the pairing interaction (note that the band index is implicitly
included in the defintion of k).  This form reduces to those considered
in Refs. 1-5 if $|k>$ is taken to be a free electron plane wave and V a
function of r-r'.  In this paper, a different approach is taken by using the
results of local density band calculations for $|k>$.  Moreover, since it is
assumed that the physics is primarily determined by the f electrons, an
approximation is made of assuming that V acts only on the J=5/2 f states on the
f atom site.  V is further approximated by asuming it has just three values
corresponding to whether the two electrons are on the same site (U), on near
neighbor sites (J), and on next near neighbor sites (JN).  Finally, the
dependence on orbital quantum number, $\mu$ (where $\mu$ ranges from -5/2 to
5/2), is taken into account by the exchange process, which is
denoted by X, with V taking the form $V^{ij}_{RR'}X_iX_j$ where i,j
represent Cartesian indices (x,y,z), R,R' denote the positions of f atom
sites, and X will be some appropriate analogue of the Pauli spin matrix,
$\sigma$.  Given this, the pairing kernel is now of the form
$$
\sum_{R,R',i,j} V^{ij}_{RR'} <k|X_i|k'>_R <-k|X_j|-k'>_{R'}
\eqno(1)
$$
where $< | >_R$ denotes an integral over a Wigner-Seitz sphere centered at R.
To evalulate this, note that the band structure wavefunction is of the form
$$
|k> = \sum_{\mu,R} a_{\mu,k,R} |\mu>_R
\eqno(2)
$$
where $|\mu>$ is the appropriate combination of spherical harmonics and spinors
times the f electron radial function.  Since X has no dependence on radial
coordinate, the radial integrals factor out of the problem, except that since
an LMTO band method is used, the wavefunction has a radial term and its
energy derivative as expansion functions.  For notational convenience, this
complication is ignored, but is taken into account when evaluating the matrix
elements.  In a band calculation, $|k>$ is just generated in one irreducible
wedge of the zone and only on the sites inside the primitive cell.  The rest
of the information for $|k>$ can be obtained by use of Bloch's theorem plus by
application of the group operations on $|k>$.  This is somewhat complicated for
$UPt_3$ since the space group is non-symmorphic with two U atoms in the
primitive cell.  This means that some group operations must
be followed by non-primitive translations and others lead to the interchange
of the two U sites.  The effect of group operations on $|\mu>$ is well known,
for $UPt_3$ they transform as $\Gamma_7, \Gamma_8,$ and $\Gamma_9$.
This has been tabulated in Ref. 7, for instance, but the author rederived
all relations used in the programs as a safety check.  Finally, in the
approximation used in this paper, Bloch's theorem will appear by factors
of the form $e^{ikR}$ in the matrix elements where R is the site position
of a U ion.  These cancel one another for the on-site case, but must be
carefully kept track of for near neighbor and next near neighbor terms.

The nature of the operator X will now be discussed.  In the $^3He$ problem,
it is taken to be the Pauli spin matrix.  The simplest generalization for the
current case is to replace $\sigma$ by the pseudo-spin matrix $\tau$, where
$\tau$ has the same effect on k, PTk (or Pk, Tk) that $\sigma$ has on up
and down spinors.  Note that P is the parity operator, and T the time reversal
operator.  Further generalizations can be made by replacing $\tau$ by
J, the total angular momentum operator.  Further, if one is interested in
quadrupolar interactions, thought by some to be of fundamental significance
in the heavy fermion problem\refto{cox}, then one can in turn replace J by
$O_2^0 = [3J_z^2-J(J+1)]/\sqrt{3}$, $O_2^1 = [J_z(J_++J_-)+(J_++J_-)J_z]/2$,
or $O_2^2 = [J_+^2+J_-^2]/2$.

To determine the appropriate symmetry of the gap function, one must project
the pairing kernel onto each group representation.  For $UPt_3$
these are $\Gamma_1$ through $\Gamma_6$, equivalent to the more
common "chemical" notation of $A_1, A_2, B_1, B_2, E_2, E_1$.  Moreover, a
projection must also be done on odd and even parity.  For even parity,
the pair function notation $|k,-k>$ is replaced by $|k,Tk> - |PTk,Pk>$.  The
odd parity case involves a three component vector.  An assumption will be made
in this case.  Experimentally, it appears from neutron scattering data
\refto{broholm} that the moments in $UPt_3$ are confined to the basal plane.
If a similar effect occurs for the Cooper pair moments, then one expects
only the "$d_z$" component of the odd parity order parameter vector to
survive.  Such an order parameter is consistent with a recent analysis
of $H_{c2}$ anisotropy in $UPt_3$\refto{choi}.  This projection automatically
occurs in the $\tau$ model if one assumes that $V^{xx}=V^{yy}$ with the rest
of the $V^{ij}$ equal to zero.  For the other models, mixing will occur.  It
is assumed, though, that there is some term in the pairing Hamiltonian which
also acts in these cases to project onto $d_z$.  For the relativistic case,
$d_z$ is $|k,Tk> + |PTk,Pk>$.

To summarize, the following pair intractions will be explored: $\tau_x\tau_x
+\tau_y\tau_y$, $J_xJ_x+J_yJ_y$, $O_2^0O_2^0$, $O_2^1O_2^1$, $O_2^2O_2^2$.
$|k,-k>$ will be taken to be $|k,Tk>\mp|PTk,Pk>$ where the upper sign is
for even parity and the lower one for $d_z$ odd parity.  The weak-coupling
gap equation to be solved is
$$
\Delta_k = \lambda^{-1} \sum_{k'} V_{k,k'} \Delta_{k'}
\eqno(3)
$$
where $V_{k,k'}$ is the pairing kernel discussed above and $\lambda^{-1}$, the
inverse coupling constant, is
$ln(1.13\omega_c/T_c)$ with $T_c$ the transition temperature and $\omega_c$
some cut-off energy.  The sum is carried out
over all k vectors on the Fermi surface.  For $UPt_3$, a 137 k vector mesh is
used on the Fermi surface in the irreducible wedge, noting that there are 24
group operations in this case.  The k vectors are weighted by the
appropriate denstiy of states factors derived from a tetrahedron decomposition
of the Brillouin zone.

In Table 1, the results of the above formalism for $UPt_3$ are summarized.
In column 2, an on-site interaction of unit value is looked at.  The sign in
each case is chosen opposite to that which would yield a nodeless $A_{1g}$
(s-wave) solution.  This sign is positive for the $\tau$, $J$, and
$O_2^2$ cases, but is negative for the $O_2^0$ and $O_2^1$ cases.  Despite
the "repulsive" value of the sign, solutions are indeed found, due to the
momentum dependence of the matrix elements.  The solution with the largest
coupling constant either has $A_{1g}$ symmetry or is odd parity (for the
$\tau$ case, the interaction is repulsive for all even parity states).
Addition
of a repulsive momentum independent constant to the gap equation did not
suppress the $A_{1g}$ solution since this solution already has
a complicated nodal structure.  In fact, the stability of the $A_{1g}$
solution over all other even parity solutions is most likely due to this
complicated nodal structure, since the nodal structure of the other
representations is fixed by symmetry and therefore not free to adjust itself
to minimize the free energy.

In the final four columns, results for ferromagnetic (positive) and
antiferromagnetic (negative) unit values of J (near neighbor) and JN (next near
neighbor) interactions are tabulated.  Again, the solution with the largest
coupling constant either has $A_{1g}$ symmetry or is odd parity.  An
interesting point to remark on is that for the  $\tau$ and $J$ cases, one
finds that the ferromagnetic sign supports odd parity solutions whereas the
antiferromagnetic sign supports even parity (for the $\tau$ case, the ferro
case is repulsive for all even parity states and the antiferro case is
repulsive
for all odd parity states).  This is in accord with earlier wisdom on this
subject\refto{miyake} but did not occur for realistic calculations on
$UPt_3$ which ignored matrix element effects\refto{norman}.  This illustrates
the crucial importance of matrix elements when evalulating the pairing kernel.
Another interesting point is that quadrupolar interactions prefer
$A_{1g}$, regardless of the sign, in ten of twelve cases.

The most favored explanation for the large body of experimental data on
$UPt_3$ is that the order parameter is from a two-dimensional group
representation with line nodes orientated perpendicular to the c axis
\refto{joynt}.  In the current context, this could be realized by $E_{1g}$
or $E_{2u}$, the latter being consistent with the $H_{c2}$ anisotropy
analysis\refto{choi}.  Neither state is favored by any of the pair
interactions analyzed in the current paper (the same was found for earlier
work which ignored matrix element effects\refto{norman}).  The most favored
two dimensional rep in the current work is $E_{1u}$.  This solution has point
nodes.  On the other hand, arguments have been put forth that the observed
line node structure in $UPt_3$ may not be due to the symmetry of the order
parameter, but rather to gaplessness due to the normal state self-energy
\refto{norm2}.  More work will certainly be necessary before the actual form
of the order parameter is unambiguously determined.

An alternate explanation of the data would be a near degeneracy of two
different representations\refto{joynt2}.  This picture is certainly supported
by the current work, which indicates several such cases which involve one
single dimensional rep and one two dimensional rep.  The picture advocated by
Machida and co-workers\refto{machida} which assumes a near degeneracy of the
three components of an odd parity solution is not consistent with the
treatment of this paper where strong spin-orbit effects prohibit such a
degeneracy.

The author concludes with some remarks on strong coupling effects.
The author has performed calculations on $UPt_3$ including the full
momentum and frequency dependence of the pairing interaction and treating
all k vectors and f bands (instead of restricting to the Fermi surface) as
recently advocated by Monthoux and Pines\refto{MPB}, but ignoring matrix
element effects.  Although this treatment leads to a renormalization of the
value of $T_c$, no change in the ordering of the solutions with respect to
coupling constant was seen.  For that reason, one would suspect that strong
coupling effects would not alter the ordering of solutions seen here, although
it is conceivable that variation of the matrix elements when going off the
Fermi surface could alter the results.  The author plans to test such
effects in the future if the effort seems warranted.  An alternate strong
coupling effect to consider is feedback to the pair potential caused by
changes induced when the metal enters the superconducting phase.  These
effects are crucial in the case of $^3He$, but for the heavy fermion case,
there has been little evidence for a change in the dynamic susceptibility
below $T_c$.  Until such changes are identified, it would be premature to
invoke such effects in the current formalism.

As a final thought, the author would like to emphasize the generality of
the above method.  Currently, this has been applied to a phenomenological and
very simple interaction vertex for $UPt_3$ as an example.  Once a
more microscopic theory is available, then the resulting calculation for
properly testing such a theory must
follow a similar formalism as expounded in this paper.  This statement will
most likely apply to other cases which might involve exchange mediated
interactions, such as the copper oxide materials.

In summary, a new method is derived for evaluating the pairing kernel in
exchange mediated superconductors utilizing matrix elements determined from
relativistic band structure calculations.  As an example, several models
for the pairing interaction for $UPt_3$ were considered,
yielding a rich variety of solutions.
Further work, both experimental and theoretical, will be necessary before
it can be determined what relevance the determined solutions have
to the experimental situation.

This work was supported by U.S. Department of Energy, Office of Basic Energy
Sciences, under Contract No. W-31-109-ENG-38.

\vfill\eject

\references

\refis{miyake} K. Miyake, S. Schmitt-Rink, and C.M. Varma, \prb 34, 6554,
1986; D.J. Scalapino, E. Loh Jr., and J.E. Hirsch, \prb 34, 8190, 1986.

\refis{norman} M.R. Norman, \prl 59, 232, 1987; \prb 37, 4987, 1988;
\prb 39, 7305, 1989; \prb 41, 170, 1990; \prb 43, 6121, 1991.

\refis{putika} W. Putikka and R. Joynt, \prb 37, 2372, 1988; \prb 39, 701,
1989.

\refis{grewe} B. Welslau and N. Grewe, Ann. Physik. {\bf 1}, 214 (1992).

\refis{MPB} P. Monthoux, A. V. Balatsky, and D. Pines, \prl 67, 3448, 1991;
A.J. Millis, \prb 45, 13047, 1992; P. Monthoux and D. Pines, \prl 69,
961, 1992; R.J. Radtke, S. Ullah, K. Levin, and M.R. Norman, to be published,
Phys. Rev. B.

\refis{and} P.W. Anderson, \prb 30, 4000, 1984.

\refis{appel} J. Appel and P. Hertel, \prb 35, 155, 1987.

\refis{cox} D.L. Cox, \prl 59, 1240, 1987.

\refis{broholm} G. Aeppli, E. Bucher, A.I. Goldman, G. Shirane, C. Broholm,
and J.K. Kjems, J. Magn. Magn. Matls. {\bf 76-77}, 385 (1988).

\refis{choi} C.H. Choi and J.A. Sauls, \prl 66, 484, 1991.

\refis{joynt} R. Joynt, J. Magn. Magn. Matls. {\bf 108}, 31 (1992) and
references therein.

\refis{norm2} M.R. Norman, Physica C {\bf 194}, 203 (1992).

\refis{joynt2} R. Joynt, V.P. Mineev, G.E. Volovik, and M.E. Zhitomirsky,
\prb 42, 2014, 1990.

\refis{machida} M. Ozaki and K. Machida, \jpsj 61, 1277, 1992.

\endreferences

\vfill\eject

\noindent Table 1.  Pairing coupling constants for various group
representations for
$UPt_3$.  U represents an on-site interaction of unit value, with the sign
chosen opposite that which would give a nodeless s-wave solution.
N and NN represent near neighbor and next near neighbor interactions of unit
value with either a positive sign (F) or negative sign (AF).  The interaction
vertex is denoted by $\tau$ for $\tau_x\tau_x+\tau_y\tau_y$, $J$ for
$J_xJ_x+J_yJ_y$, $O_2^0$ for $O_2^0O_2^0$, $O_2^1$ for $O_2^1O_2^1$, and
$O_2^2$ for $O_2^2O_2^2$.  Listed are the group representations with the
largest coupling constants (with the coupling constant in parenthesis).

\settabs 6 \columns
\vskip12pt
\hrule
\vskip6pt
\+Case&U&F-N&AF-N&F-NN&AF-NN\cr
\vskip6pt
\hrule
\vskip6pt
\+$\tau$&$B_{1u}$(0.015)&$E_{1u}$(0.040)&$A_{1g}$(0.064)&$B_{1u}$(0.047)&
$A_{1g}$(0.164)\cr
\+&$A_{2u}$(0.009)&$B_{1u}$(0.036)&$E_{2g}$(0.047)&$E_{1u}$(0.046)&
$E_{2g}$(0.070)\cr
\vskip6pt
\hrule
\vskip6pt
\+$J$&$A_{1g}$(0.058)&$E_{1u}$(0.188)&$A_{1g}$(0.209)&$B_{1u}$(0.280)&
$A_{1g}$(0.728)\cr
\+&$B_{1u}$(0.047)&$B_{1u}$(0.169)&$E_{2g}$(0.172)&$E_{1u}$(0.236)&
$E_{2g}$(0.397)\cr
\+&&&&$A_{1g}$(0.207)\cr
\vskip6pt
\hrule
\vskip6pt
\+$O_2^0$&$A_{2u}$(0.247)&$A_{1g}$(0.838)&$A_{1g}$(1.33 )&$A_{1g}$(2.32 )&
$A_{2u}$(0.484)\cr
\+&$E_{1u}$(0.171)&$A_{2u}$(0.699)&$E_{2g}$(0.600)&$E_{1u}$(0.635)&
$B_{1u}$(0.334)\cr
\+&$E_{2g}$(0.161)&$E_{1u}$(0.392)&$E_{1u}$(0.413)&$B_{1u}$(0.629)&
$A_{1g}$(0.328)\cr
\+&&&&&$E_{1u}$(0.323)\cr
\vskip6pt
\hrule
\vskip6pt
\+$O_2^1$&$A_{1g}$(0.157)&$A_{1g}$(0.438)&$A_{1g}$(0.361)&$A_{1g}$(1.60 )&
$A_{1g}$(0.547)\cr
\+&$B_{1u}$(0.121)&$A_{2u}$(0.216)&$E_{2g}$(0.227)&$E_{2g}$(0.486)&
$E_{1u}$(0.323)\cr
\+&$E_{2g}$(0.102)&$B_{1u}$(0.215)&$A_{2u}$(0.186)&&$B_{1u}$(0.272)\cr
\+&&$E_{1u}$(0.194)\cr
\vskip6pt
\hrule
\vskip6pt
\+$O_2^2$&$A_{1g}$(0.145)&$A_{1g}$(0.374)&$A_{1g}$(0.236)&$A_{1g}$(0.381)&
$A_{1g}$(1.05 )\cr
\+&$B_{1u}$(0.141)&$E_{1g}$(0.100)&$E_{1u}$(0.235)&$E_{2g}$(0.174)&
$E_{2g}$(0.571)\cr
\+&&&$E_{2g}$(0.166)&$B_{1u}$(0.163)&$B_{1u}$(0.392)\cr
\vskip6pt
\hrule

\vfill\eject

\endit